\documentclass{PoS} 

\title{Deconfining phase transition on a double-layered torus}

\ShortTitle{Deconfining phase transition on a double-layered torus}

\author{\speaker{Bernd A. Berg}\\ Department of Physics,
        Florida State University, Tallahassee, FL 32306, USA}

\author{Alexei Bazavov\\ Department of Physics, 
        University of Arizona, Tucson, AZ 85721, USA}

\author{Hao Wu\\ Department of Physics, 
        Florida State University, Tallahassee, FL 32306, USA}

\abstract{Deconfined regions in relativistic heavy ion collisions are 
limited to small volumes surrounded by a confined exterior. Here the 
geometry of a double layered torus is discussed, which allows for 
different temperatures in its two layers. This geometry enables one 
to approach the QCD continuum limit for small deconfined volumes with 
confined exteriors in a more realistic fashion than by using periodic 
boundary conditions.  Preliminary data from a study for pure SU(3) 
lattice gauge theory support a substantial increase in a pseudo 
transition temperature.}

\FullConference{The XXVII International Symposium on Lattice Field Theory\\
                 July 26-31, 2009\\
                 Peking University, Beijing, China}

\begin{document}

\section{Equilibrium simulations of the deconfinement transition and lab
plasma}

Statistical properties of a quantum system with Hamiltonian $H$ in
a continuum volume $V$, which is in equilibrium with a heatbath 
at physical temperature $T$, are determined by the {\it partition 
function}
\begin{equation}
    Z(T,V)={\rm Tr} e^{-H/T}=
    \sum_{\phi}\langle\phi|e^{-H/T}|\phi\rangle,
\end{equation}
where the sum extends over all states and the Boltzmann constant is 
set to one. Imposing periodic boundary conditions (PBC) in Euclidean 
time $\tau$ and bounds of integration from $0$ to $1/T$, one can 
rewrite the partition function in the {\it path integral representation:}
\begin{equation}
    Z(T,V)=\int D\phi\exp\left\{-\int_0^{1/T}d\tau L_E
    (\phi,\dot\phi)\right\}.
\end{equation}
Nothing in this formulation requires to carry out the infinite
volume limit. 

Past LGT simulations of the deconfining transition focused primarily 
on boundary conditions (BC), which are favorable for reaching the 
infinite volume quantum continuum limit (thermodynamic limit of the 
textbooks) quickly with temperature and volume of the system given by
\begin{equation} 
  T = \frac{1}{a\,N_{\tau}} = \frac{1}{L_{\tau}}\,,~~~N_s/N_{\tau}\to
  \infty\,,~~N_{\tau}\to\infty\,,~~L_{\tau}~{\rm finite}\,,
\end{equation}
where $a$ is the lattice spacing. These are {\it PBC} in the spatial 
volume $V = (a\,N_s)^3$. For the deconfinement phase created in a 
heavy ion collision the infinite volume limit does not apply. Instead 
we have to take the {\it finite volume continuum limit}
\begin{equation} 
  N_s/N_{\tau} = {\rm finite}\,,~~N_{\tau}\to\infty\,,~~L_{\tau}~
  {\rm finite}\,,
\end{equation}
{\it and PBC are incorrect} because the outside is in the 
confined phase at low temperature. E.g., at the BNL RHIC one expects 
to create an {\it ensemble of differently shaped and sized deconfined 
volumes.} The largest volumes are those encountered in central 
collisions. A rough estimate of their size is
\begin{eqnarray} 
  &\pi\times (0.6\times {\rm Au\ radius})^2\times c\times
  ({\rm expansion\ time}) & \\ \nonumber
  &= (55\ {\rm fermi}^2)\times ({\rm a\ few\ fermi}) &
\end{eqnarray}
where $c$ is the speed of light. Here we want to estimate 
finite volume corrections for pure SU(3) and focus 
on the continuum limit for
\begin{equation} 
  L_s = a N_s = (5-10)\ {\rm fermi}\ .
\end{equation}
In the following we set the physical scale by 
\begin{equation} 
  T^c=174\ {\rm MeV}\ ,
\end{equation}
which is in the range of QCD estimates with two light flavor quarks, 
implying for the temporal extension \begin{equation} 
  L_{\tau} = a\,N_{\tau} = 1.13\ {\rm fermi}\ .
\end{equation}

\section{Simulations with cold boundary conditions}

We consider difficulties and effects encountered when one equilibrates 
a hot volume with cold boundaries by means of Monte Carlo (MC) 
simulations for which the updating process provides the equilibrium. 
We use the single plaquette {\it Wilson action} on a 4D hypercubic 
lattice. Numerical evidence shows that SU(3) lattice gauge theory 
exhibits a weakly first-order deconfining phase transition at some 
coupling $\beta^g_t(N_{\tau})=6/g^2_t (N_{\tau})$. {\it The scaling 
behavior} of the deconfining temperature is
\begin{equation} 
  T^c = c_T\,\Lambda_L
\end{equation}
where the lambda lattice scale
\begin{equation} 
  a\,\Lambda_L = f_{\lambda}(\beta^g) = \lambda(g^2)\,
  \left(b_0\,g^2\right)^{-b_1/(2b_0^2)}\,e^{-1/(2b_0\,g^2)}\,,
\end{equation}
has been {\it determined in the literature}. The coefficients $b_0$ 
and $b_1$ are perturbatively obtained from the renormalization group 
equation,
\begin{equation}
  b_0 = \frac{11}{3}\frac{3}{16\pi^2}~~{\rm and}~~
  b_1=\frac{34}{3}\left(\frac{3}{16\pi^2}\right)^2\,.
\end{equation}
Higher perturbative and non-perturbative corrections are parametrized 
in \cite{BBV06} by
\begin{equation} 
  \lambda(g^2)\ =\ 1+a_1\,e^{-a_2/g^2}+a_3\,g^2+a_4\,g^4
  ~~~{\rm with}
\end{equation}
$a_1=71553750,\ a_2=19.48099,\ a_3=-0.03772473,\ a_4=0.5089052.$
In the region accessible by MC simulations this parametrization is 
perfectly consistent with an independent earlier one in \cite{NeSo02},
but has the advantage to reduce for $g^2\to 0$ to the perturbative 
limit.

Imagine an almost infinite space volume $V=L_s^3$ and a smaller 
sub-volume $V_1=L_{s,1}^3.$ The complement to $V_1$ in $V$ will be 
called $V_0$ (outside world). The number of temporal lattice links 
$N_{\tau}$ is the same for both volumes. We like to find parameters 
so that scaling holds in both volumes, while $V_1$ is at temperature 
$T_1=T_c$ and $V_0$ at room temperature $T_0$.
We denote the coupling by $\beta^g_0$ for plaquettes in $V_0$ and by 
$\beta^g_1$ for plaquettes in $V_1$. For that purpose any plaquette 
touching a site in $V_1$ is considered to be in $V_1$. We would like to 
have $\beta^g_1$ in the scaling region, say $\beta^g_1=6.$ The relation
\begin{equation}
  10^{10}\approx\frac{T_1}{T_0} = \frac{a_0}{a_1} = 
  \frac{f_{\lambda}(\beta_0^g)\,\Lambda_L}{f_{\lambda}(\beta^g_1)
  \,\Lambda_L} = \frac{f_{\lambda}(\beta_0^g)}{f_{\lambda}(\beta^g_1)}
\end{equation}
drives $\beta^g_0$ out of the scaling region and (after moving over
to strong coupling relations) practically to $\beta^g_0=0$, which
we call {\it disorder wall} BC. In the disorder wall approximation of 
the cold exterior we can simply omit contributions from plaquettes 
that involve links through the boundary. MC simulations, which we
quickly summarize here, were performed in Ref.\cite{BaBe07}. Due 
to the use of the strong coupling limit for the outside volume, 
{\it scaling of the results is not obvious.} 

We use the maxima of the Polyakov loop susceptibility
\begin{equation}
  \chi_{\max}=\frac{1}{N_s^3}\left[\langle|P|^2\rangle-
  \langle|P|\rangle^2\right]_{\max},\,\,\,P=\sum_{\vec{x}}P_{\vec{x}}
\end{equation}
to define pseudo-transition couplings $\beta^g_{pt}(N_s;N_{\tau})$.
Our BC introduce an order $N_s^2$ disturbance, so that 
\begin{equation} 
  \beta^g_{pt}(N_s;N_{\tau}) = \beta^g_t(N_{\tau}) +
  a_1^d\,\frac{N_{\tau}}{N_s} +
  a_2^d\,\left(\frac{N_{\tau}}{N_s}\right)^2 +
  a_3^d\,\left(\frac{N_{\tau}}{N_s}\right)^3 +\ \dots
\end{equation}
holds.
\begin{figure} \begin{center}
\includegraphics[width=0.7\textwidth]{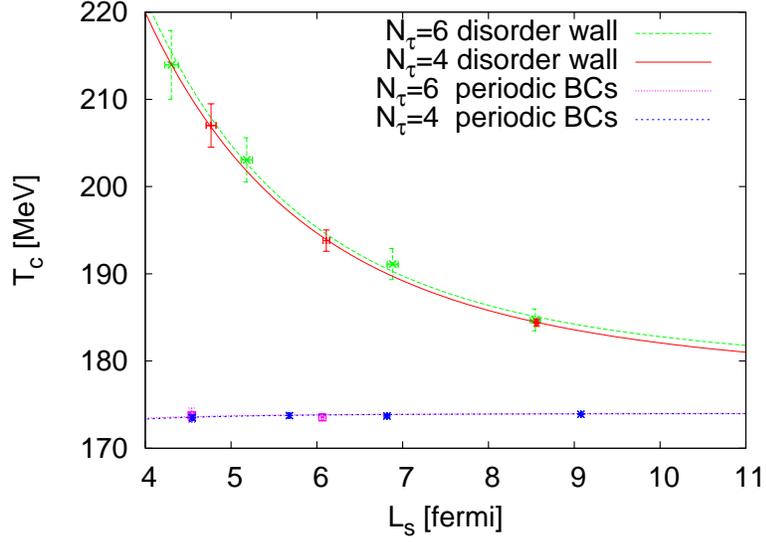} 
\caption{Estimates of finite volume corrections to $T_c$.} 
\label{fig_Tc_Ls}
\end{center} \end{figure} 
\begin{figure} \begin{center}
\includegraphics[width=0.7\textwidth]{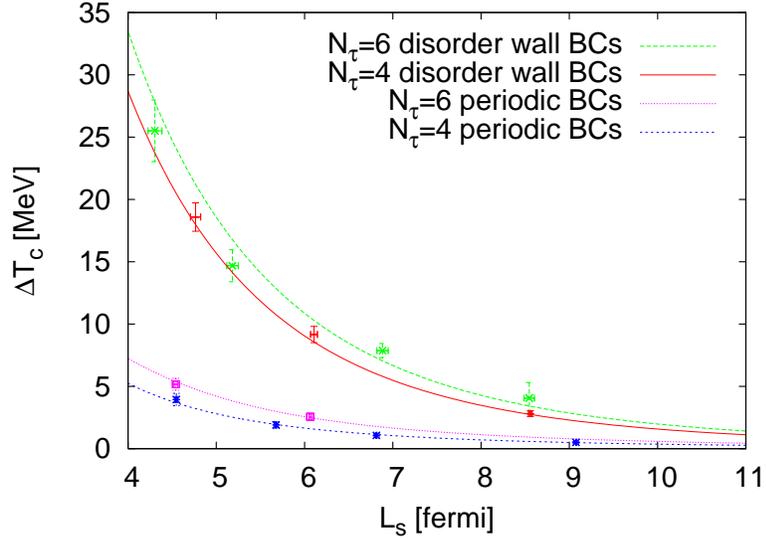} 
\caption{Estimates of finite volume corrections to the width of the
Polyakov loop susceptibility.} \label{fig_width}
\end{center} \end{figure} 
Fits of pseudo-transition coupling constant values lead to
estimates of finite volume corrections to $T_c$ as given in 
Fig.~\ref{fig_Tc_Ls}. It is seen from the figure that they are
consistent with scaling as the curves for $N_{\tau}=4$ and
$N_{\tau}=6$ fall on top of one another. There are {\it no free 
parameters involved} at this point, because  the non-perturbative
parametrization of the SU(3) lambda scale was determined in
independent literature. Estimates of finite volume corrections 
to the full width of Polyakov loop susceptibilities at 2/3 maximum 
are shown in Fig.~\ref{fig_width}.  Again, the curves are found to 
be consistent with scaling.

Results so far show that for volumes of BNL RHIC size the magnitudes 
of $T_c$ and width corrections are comparable to those obtained by 
including quarks into pure SU(3) calculations ($T_c$ in the opposite, 
width in the same direction). Similar corrections are expected for the 
equation of state. Previous QCD calculations at finite temperatures 
and densities should therefore be extended to cold BC.

However, there remain problematic questions about disorder wall BC.
Although the results show scaling, it is unsatisfactory that the 
disorder wall BC do not reflect a physical outside volume. Two 
properties are desirable:
\begin{enumerate}
\item Inside and outside volumes are kept in the scaling region.
\item The spatial lattice spacing $a_s$ is the same on both sides 
      of the boundary. 
\end{enumerate}
In such simulations one may again keep $T_{\rm inside}$ at $T_c$ and
study its dependence on $T_{\rm outside}$. As outlined in the next 
section, this can be done in the newly introduced geometry \cite{Be09} 
of a double layered torus (DLT), but has then to be done for a rather 
small $T_{\rm outside}$ interval
\begin{equation}
  T_{\rm outside}\in \left[T_c-\triangle T,T_c\right]\,,
  ~~\triangle T>0\,.
\end{equation}

\section{Simulations on a double layered torus}

\begin{figure} \begin{center}
\includegraphics[width=0.7\textwidth]{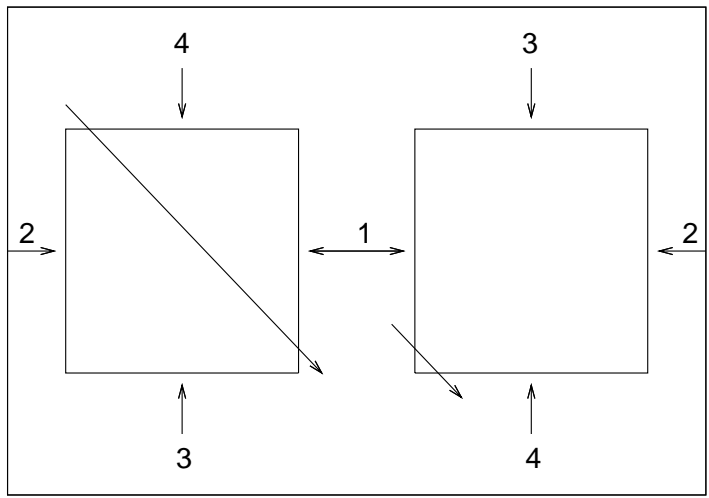} 
\caption{Double layered torus in two dimensions.} \label{fig_DLT}
\end{center} \end{figure} 

For the DLT the boundaries are glued together as indicated by the 
arrows in Fig.~\ref{fig_DLT}. Note that interchanging labels 3 and 4 
on one of the lattices leads to a situation in which some sites are 
connected by two links and the different geometry of a sphere.

With DLT BC in the spacelike directions and PBC in the fourth
direction one can simulate at two temperatures and each volume 
becomes the outside world of the other. The two temperatures are 
adjusted by tuning the coupling constants in the two volumes. MPI 
Fortran code for SU(3) simulations on a DLT is given in 
Ref.\cite{Be09,BeWu09}. Preliminary numerical results with both
temperatures in the SU(3) scaling region are compiled in the following.

\begin{figure} \begin{center} 
\includegraphics[width=0.7\textwidth]{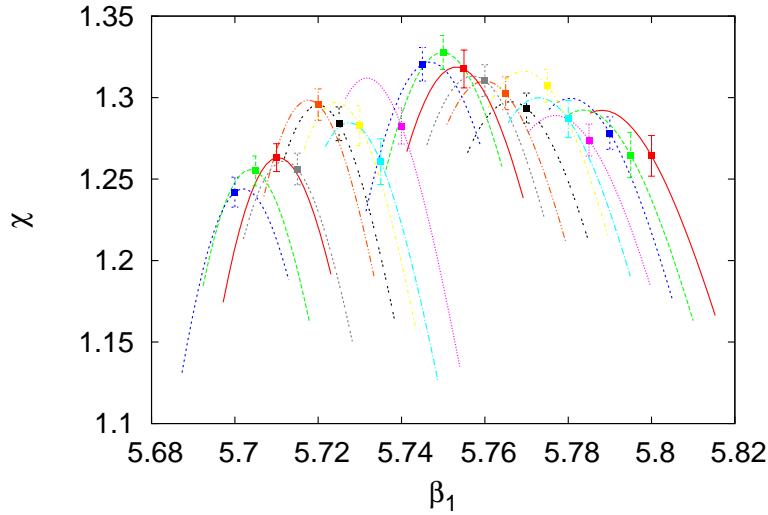} 
\caption{Polyakov loop susceptibilities on a $8^34$ DLT.} 
\label{fig_CPm}
\end{center} \end{figure}

Figure~\ref{fig_CPm} shows the reweighting in $\beta_1$ of Polyakov 
loop susceptibilities on a $8^34$ lattice, each simulation point
corresponding to a different pair of coupling constant values
$(\beta_0^0,\beta_1^0)$, adjusted so that $\beta_1^0$ is close
to the pseudocritical point. 

\begin{figure} \begin{center} 
\includegraphics[width=0.7\textwidth]{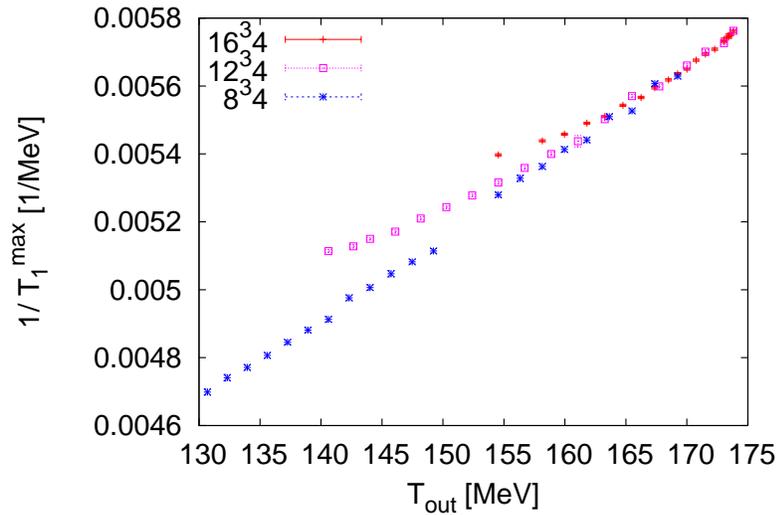} 
\caption{Inverse pseudo transition temperatures versus outside
temperatures.} \label{fig_1oTcDat}
\end{center} \end{figure}

\begin{figure} \begin{center} 
\includegraphics[width=0.7\textwidth]{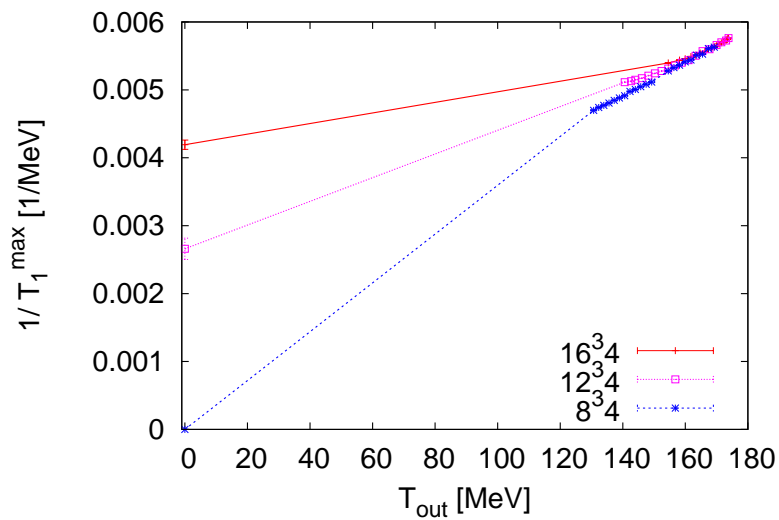} 
\caption{Results of the previous figure together with disorder wall 
estimates.} \label{fig_1oTc} \end{center} \end{figure} 

Using scaling relations the inverse physical pseudocritical 
temperatures, $1/T_1^{\max}$ from all our lattice sizes, with 
$T_1^{\max}$ corresponding to the maxima of the Polyakov loop 
susceptibilities, are plotted in Fig.~\ref{fig_1oTcDat} versus the 
outside temperature $T_{\rm out}$. Even for the small range of 
outside temperatures in the SU(3) scaling region, one sees already 
sizable corrections of $T_1^{\max}$. The same data are plotted in 
Fig.~\ref{fig_1oTc}, extending the $T_{\rm out}$ down to zero, so 
that the estimates from Ref.\cite{BaBe07} can be included (on
the $8^34$ lattice not transition was found with disorder wall
BC as indicated here by $1/T_1^{\max}=0$).

In summary, our preliminary DLT results in the scaling region are 
consistent with the disorder wall finite volume $T_c$ estimates.
These simulations keep inside and outside temperatures in the SU(3)
scaling region. To achieve also a continuous spacelike lattice spacing
across the boundary, one has to use asymmetric couplings in space
and time directions. Considerable work remain to be done to obtain 
an overall convincing description.

{\bf Acknowledgments:} This work has in part been supported by DOE 
grants DE-FG02-97ER-41022, DE-FC02-06ER-41439, by NSF grant 0555397,
and by the German Humboldt Foundation. BB thanks Wolfhard Janke and 
his group for their kind hospitality at Leipzig University. Simulations 
were performed at NERSC under ERCAP 82860 and on PC clusters at FSU.

\end{document}